\documentclass[11pt,a4paper,twocolumn,final]{article}
\usepackage[utf8]{inputenc}
\usepackage[english]{babel}
\usepackage[left=2cm,right=2cm,top=2cm,bottom=2cm]{geometry}

\usepackage[affil-it]{authblk}
\usepackage{comment}
\usepackage{amsmath,tabu,amssymb,amsfonts,amsmath,varwidth,graphicx,color,xcolor,subcaption}
\usepackage{caption}
\usepackage{epstopdf}

\newcommand{\DE}{\mathrm{d}}

\newcommand{\dder}[2]{\frac{\DE^2#1}{\DE#2^2}}

\newcommand{\dinteg}[4]{\int_{#1}^{#2}\!{#3}\,\DE #4}

\begin{document}
\title{Influence of fracture criteria on dynamic fracture propagation in a discrete chain}

\author[1]{Nikolai Gorbushin}
\author[1]{Gennaro Vitucci}
\affil[1]{Department of Mathematics, IMPACS, Aberystwyth University, UK}
\author[2]{Grigory Volkov}
\affil[2]{Department of Theory of Elasticity, St. Petersburg State University, Russia}
\author[1]{Gennady Mishuris}
\date{\vspace{-5ex}}
\maketitle

\section*{Abstract}
The extent to which time-dependent fracture criteria affect the dynamic behavior of fracture in a discrete structure is discussed in this work. The simplest case of a semi-infinite isotropic chain of oscillators has been studied. Two history-dependent criteria are compared to the classical one of threshold elongation for linear bonds. The results show that steady-state regimes can be reached in the low subsonic crack speed range where it is impossible according to the classical criterion. Repercussions in terms of load and crack opening versus velocity are explained in detail. A strong qualitative influence of history-dependent criteria is observed at low subsonic crack velocities, especially in relation to achievable steady-state propagation regimes.

\section{Introduction}
Studying fracture propagating in discrete structures results in a tool capable of analyzing a broad range of phenomena which would not emerge in the settings of continuum mechanics. The approach has found fruitful applications when dealing with crystals, cellular materials, cracks in fiber-reinforced matrices and investigations at the atomic level (e.g.\cite{abraham1998spanning,atrash2011evaluation,bolander2005irregular,tsai2010characterizing,glassmaker2007biologically}). Lattice structure models become even more crucial in the framework of dynamic propagation. With this respect, many kinds of instabilities can be predicted by intuitive considerations without the need of ad hoc hypotheses (see \cite{marder1995origin,bernstein2003lattice,kessler2003does,Bitzek2015}). Finally, discrete models can be also treated as the discretization of the corresponding continuum problems
where also a choice of the fracture criterion may play an important role when dynamic fracture propagation is in question, for example it may lead
to different predictions on the stability of possible steady-state regimes (e.g. \cite{Morini2013,Mishuris_Picci2012}).

A lattice structure, in the dynamic scenario, is composed of concentrated masses interacting via links characterized by an interaction potential. In this paper the latter is parabolic, relating only the closest neighbors, while non-local interactions have been studied in \cite{gorbushin2016analysis}. The analyzed structure is mono-dimensional: a chain of oscillators, which is detached from a substrate that reflects the problem symmetry.

The focus of this work is to investigate the influence of the fracture criteria of the links on the dynamic fracture propagation in such medium. {Exclusively cracks which advance at constant speed, in a steady-state, are analyzed. Such regimes, indeed, have traditionally been of extreme interest in the field of dynamic fracture and have repeatedly been observed experimentally. A few classical studies can be highlighted for example in \cite{ravi1984experimental,fineberg1991instability,hauch1999dynamic} while the topic gained new attention more recently in discrete structures such as bridges \cite{brun2014transition} or xyloexplosives \cite{stick09}.}{}

Before addressing the problem of the propagation, though, the behavior of a single link is discussed hereafter. A linear elastic spring can be quasi-statically elongated to failure and its final elongation value, $u_s$, supposed to be known constant. The simplest and most common failure criterion neglects dynamic effects on the spring resistance. It identifies the displacement $u$ as critical when
\begin{equation}
\min{t}:\quad u(t)=u_s,
\label{eq:idbrit0}
\end{equation}
by which the time $t=t_f$ when the fracture occurs can be found.

A fracture event, though, in many materials turns out to be not simply determined by an instantaneous threshold value for some energy measure like the maximum elongation established above. We deal in the present work with non-instantaneous fracture criteria which nevertheless do not change the material stiffness. The rate with which a body is deformed or an integral measure of the deformation energy provided to a bond before it breaks are examined: the incubation time (IT) and the Tuler-Butcher (TB) criteria.

According to the first of those formulations, the average stress, or equivalently the average linear elastic stretch, over a period of time preceding the breakage is considered as the cause of fracture. Such a period is actually called the incubation time $\tau$. The criterion, originally formulated in terms of stresses in \cite{Petrov_Utkin89}, can be written for the elastic bond in object in terms of the elongation as
\begin{equation}
\min{t}:\quad \frac{1}{\tau}\dinteg{t-\tau}{t}{u(\theta)}{\theta}=u_s.
\label{eq:itor0}
\end{equation}
Notice that $u_s$ is still the threshold elongation of the spring when measured statically.
The main idea of this approach is that a transient process does not occur instantly but includes a more complicated breakage process. Its realization demands a time period depending on the intensity and shape of loading and on the internal structure of the fractured media. The introduction of the incubation time $\tau$ as strength parameter in addition to $u_s$ allows one to predict the stress level at the instant of fracture for a variety of loading pulses with different intensities and shapes (\cite{volkov2017some}). So the physical meaning of the fracture incubation time is a characteristic time determining the material's ability to resist to dynamic loading. The IT approach has shown to be reliable in different branches of mechanics and physics, such as the dynamic fracture of rocks and concretes, the dynamic yielding of metals, the acoustic ultrasonic cavitation of liquids, etc (\cite{Petrov2004}; \cite{Gruzdkov_Petrov}). Particularly, this criterion was successfully applied to problems of fracture in materials with pre-existing cracks. In this case the criterion can be reformulated in terms of stress intensity factor. The steady-state fracture propagation in a lattice structure is indeed crack growth and this analogy allows the application of the criterion to the problem considered here.

On the other hand, it has been observed that cumulative damage can also be the cause of fracture. A way to quantify it is via the Tuler-Butcher criterion discussed in \cite{tuler1968criterion}. Again, one linear elastic bond is statically tested until it breaks at $u=u_s$. In a TB material, $u_s$ represents the elongation to be exceeded in dynamics as
\begin{equation}
\min{t}:\quad \dinteg{0}{t}{\mathrm{H}(u(\theta)-u_s)\left(\frac{u(\theta)}{u_s}-1\right)^2  }{\theta}=D
\label{eq:TBcrit0}
\end{equation}
before the fracture occurs at $t=t_f$. Here, $\mathrm{H}(u-u_s)$ is the Heaviside step function by which it is possible to write that only the work of the overstretch $u-u_s$ contributes to damage. Note that also this criterion was originally formulated in terms of stresses and that the exponent two was left general in the original formulation, but turns out to be such in most experiments. In this way the physical meaning of the criterion is that a maximum work has to be done by an external overload on the spring before it collapses. Looking at Eq.\eqref{eq:TBcrit0}, it turns out that, as for the IT criterion, TB materials can be regarded as one possible extension of ideal brittleness. The latter can be retrieved indeed by setting the cumulated energetic damage $D$ to zero. The criterion has found fruitful applications in analyzing spallation, impact loading, thermal shock caused fracture in rocks, glass, aluminum, copper (see \cite{boustie1991experimental}; \cite{wei2009modeling}; \cite{grady2010length}).

In order to illustrate the peculiarities of the criteria, one can imagine applying a ramp displacement of rate $r$ at $t=0$ to three links with the same static strength $u_s$ but different failure behavior. The ideally brittle one would break at $t_f=u_s/r$ as soon as its elongation reaches $u_s$. The IT spring would break, accordingly to Eq.\eqref{eq:itor0}, at $t_f=u_s/r+\tau/2$ if $r<2u_s/\tau$ or at $t_f=\sqrt{2u_s\tau/r}$ otherwise, thus establishing a distinction between low and high deformation rates. The criterion shows a delayed failure causing an ultimate elongation bigger than the static one in this loading condition. In the case of a non-monotonic load, though, such delay might result in an elongation at failure which is smaller than the static one or during unloading (see \cite{volkov2017some}).
The TB criterion also predicts a delayed failure, but at $t_f=u_s/r+u_s\sqrt[3]{3D^2/r^2}$ according to \eqref{eq:TBcrit0}. The difference with the IT case is that, now, an oscillating load which is strong enough to break the spring in statics will also do it in dynamics. Notice in fact that a constantly increasing cumulated damage would sooner or later surpass $D$ (see left-hand side in Eq.\eqref{eq:TBcrit0}).

The rest of the paper is devoted to the model of a fracture in a structured medium subjected to the aforementioned criteria and their effects on the stable regimes of propagation. The results are expressed in terms of trapped lattice energy, applied remote force and crack tip opening.

\begin{figure}[!ht]
\captionsetup{justification=raggedright}
\centering
\includegraphics[width=0.45\textwidth]{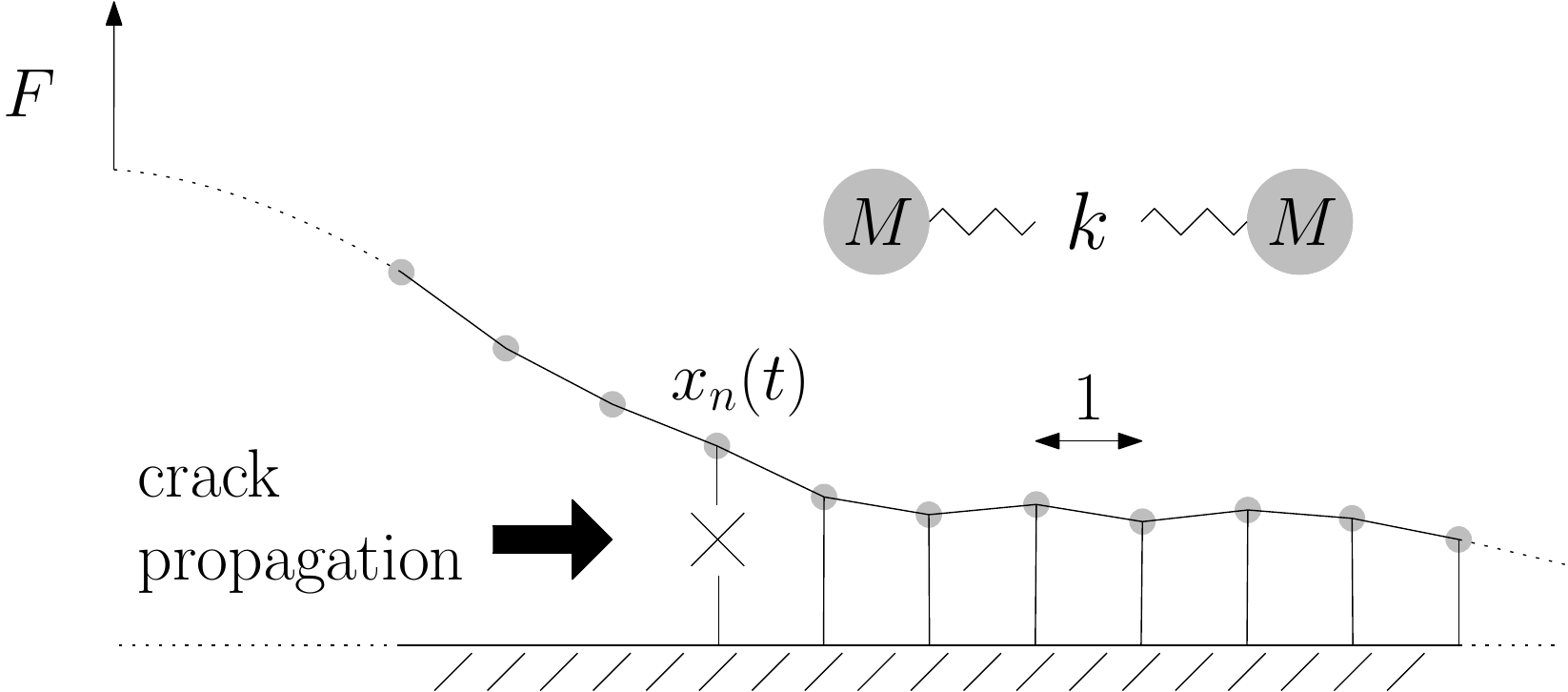}
\caption{Crack propagation in an infinite discrete chain. An infinite number of masses $M$ separated by a unitary distance are linked to each other and to a rigid substrate via linear elastic springs of stiffness $k$. A force $F$ applied at a remote distance breaks the links and the crack propagates to the right. }
\label{fig:chain2}
\end{figure}

\section{Background}\label{sec:back}
Consider an infinite number of masses $M$ linked to each other and to a rigid substrate via linear elastic springs of length 1 and stiffness $k$. A force $F$, applied at infinite distance, introduces energy into the system and finally breaks the links causing the crack to propagate to the right in Fig.~\ref{fig:chain2}. If $x_n(t)$ is the link that fails at the time $t$, the equilibrium for a generic oscillator at $x=x_i$ in terms of its displacement $u_i(t)$ holds as
\begin{equation}
\frac{M}{k}\dder{u_i}{t}=u_{i+1}+u_{i-1}-\left(2+H(i-n)\right)u_i,
\label{eq:dorig}
\end{equation}
where the Heaviside step function $H(i-n)$ allows for the combination of the equation for the detached ($i<n$) and intact part ($i\geq n$) of the chain.

Only fracture with a constant speed $v$, i.e. steady-state fracture, is analyzed here. In such a way the problem is reduced to the long known settings of \cite{slepyan1984fracture,slepyan2012models}. The fracture can travel slower than sound in the broken structure, that is
\begin{equation}
v<v_c\text{,}\qquad v_c=\sqrt{\frac{k}{M}}.
\end{equation}
As a result of the steady-state assumption we search for a solution in terms of the unknown function
\begin{equation}
u(x_i-vt)=u_i(t),
\label{eq:eqidef}
\end{equation}
for any $i$ and $t>0$.

Further on we adopt a coordinate system which moves together with the crack tip
\begin{equation}
\eta=x-x_n(t)=x-vt,
\label{eq:teta}
\end{equation}
in a way that the crack tip sits conveniently always at $\eta=0$. By using such new moving frame, the coordinate $\eta$ accounts for time and position simultaneously. Thus, the equation of motion Eq.\eqref{eq:dorig} for $u(\eta)$ can be written in the broken ($\eta<0$) and intact ($\eta\geq 0$) sides of the tip as:
\begin{equation}
\dder{u(\eta)}{\eta}=\frac{u(\eta+1)+u(\eta-1)-\left(2+H(\eta)\right)u(\eta)}{v^2/v_c^2}.
\label{eq:orig}
\end{equation}

With the help of the mathematical tools of Fourier transform and Wiener-Hopf technique, such an equation has been repeatedly solved for this and more complex structures, for example in \cite{slepyan1999lattice,marder1995origin,kresse2004lattice,nieves2016analysis,gorbushin2016analysis}, such to give the displacement profile $u(\eta)$ which travels along the structure at a given steady-state crack speed $v$. If one intends to describe the trajectory of a single mass during time, one has to just apply Eq.\eqref{eq:eqidef}.

\begin{figure}[ht!] 
\captionsetup{justification=raggedright}
  \begin{subfigure}[b]{.9\linewidth}
    \centering
    \includegraphics[width=\linewidth]{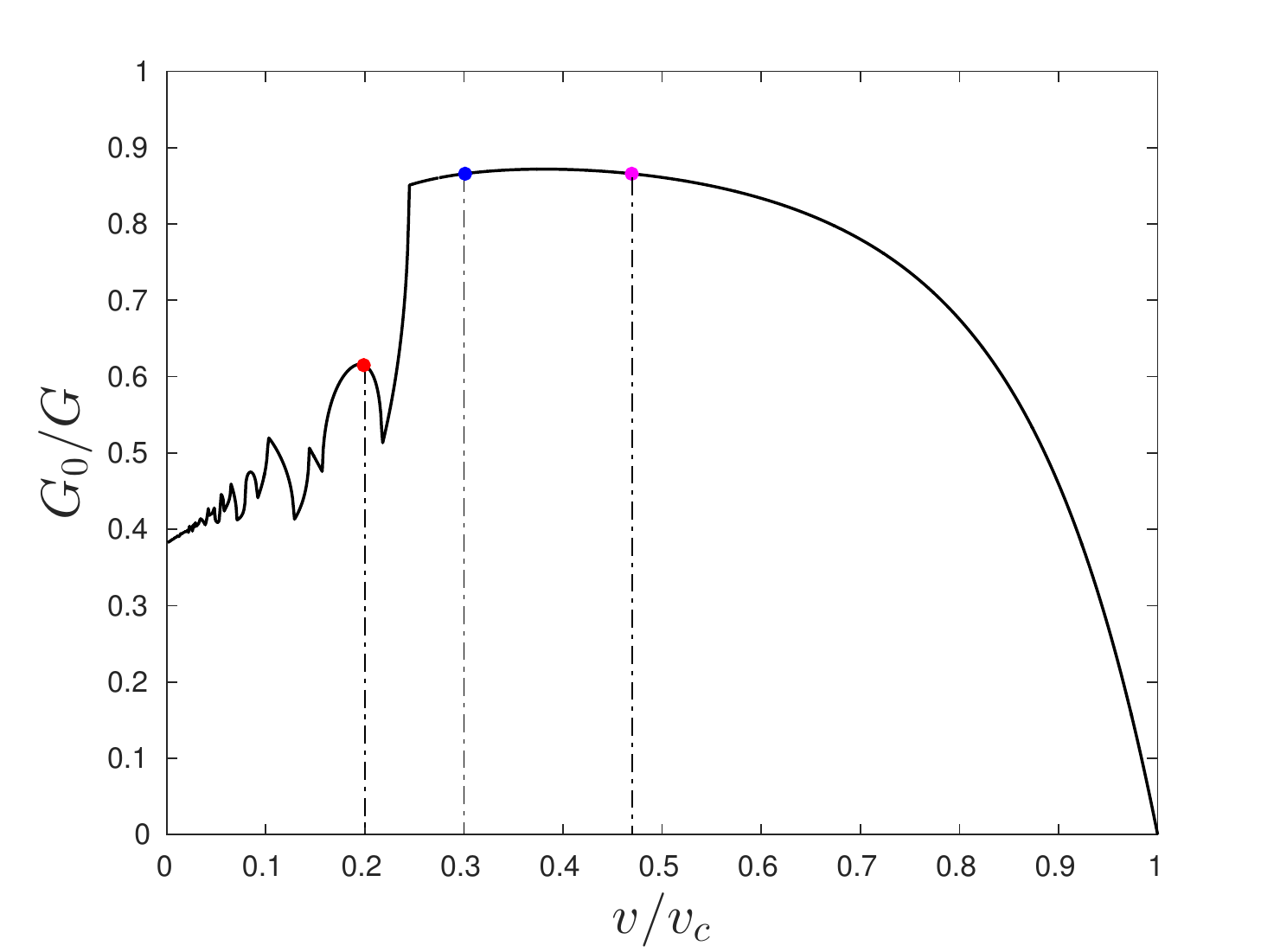} 
    \caption{Energy release rate.}  
    \label{fig:err1} 
    \centering
    \includegraphics[width=\linewidth]{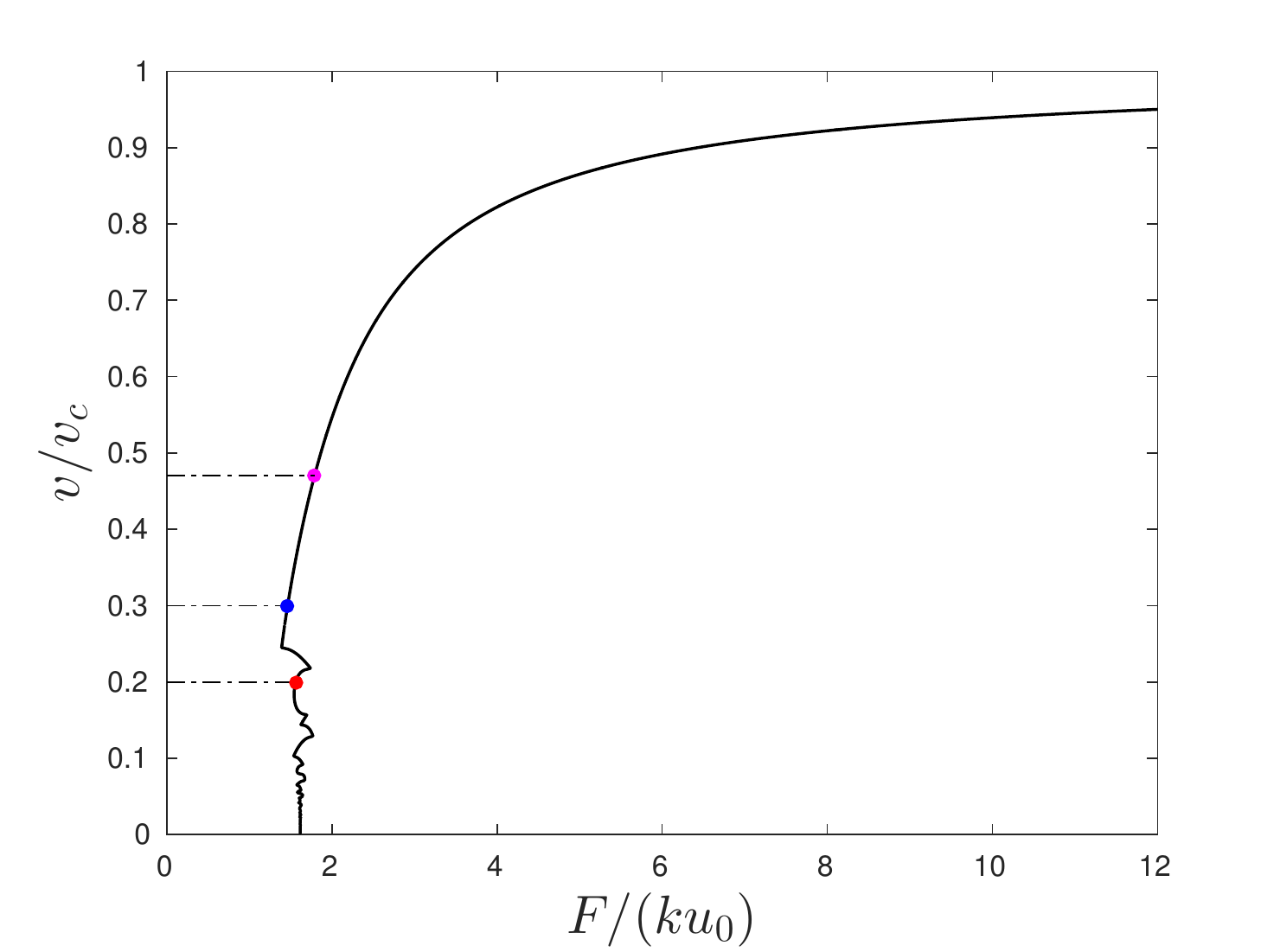} 
    \caption{Force versus crack speed.} 
    \label{fig:fvs1} 
  \end{subfigure}
  \caption{Existence of steady state solutions for the fracture propagation problem in a discrete chain as in \cite{slepyan1984fracture}. The dimensionless energy release rate and remotely applied force are plotted as functions of the normalized crack speed. For the displacement profiles of the points highlighted on the curves see Fig.\ref{fig:dispb}.}
  \label{fig:preq}
\end{figure}

The concept of energetic lattice trapping of a structured material was originally introduced in \cite{thomson1971lattice} and it can be quantified by the ratio $G/G_0$ not smaller than unity. Such energy may be introduced into the system in different ways. Analytical relations for the energy release rate $G$ for every crack velocity have been retrieved in \cite{slepyan1984fracture}. The quantity
\begin{equation}
G_0=\frac{k u_0^2}{2}
\end{equation}
is the link strain energy which is released locally at the crack tip at the moment of fracture, where $u_0=u(0)$. For the examined chain, $G_0/G$ has been plotted in Fig.~\ref{fig:err1}. In our work we assume that the energy derives from a constant force $F$ far away and the consequent crack speed has been recently derived in \cite{gorbushin2017dynamic} as
\begin{equation}
\frac{F}{k u_0}=\sqrt{2\frac{G}{G_0}\frac{v_c+v}{v_c-v}}:=\Phi(v)
\label{eq:fove0}
\end{equation}
and plotted in Fig.~\ref{fig:fvs1}. {Note that the same curve also implies that the limiting velocity $v_c$ can not be reached via a finite force besides requiring an infinite energy release rate (as from Fig.~\ref{fig:err1}).}{}

The assumption that the crack propagates at a constant speed also requires some additional consideration. In particular it means, for a given oscillator $i$ sitting at $x_i$, that it is not allowed to break before all the links situated at $x<x_i$ do (links on the left-hand side of Fig.\ref{fig:chain2}). From the propagation point of view, it must be clarified that a regime which involves nucleation of daughter cracks ahead of the mother crack tip ($\eta>0$) is non-admissible. The detachment of the chain has to progress continuously. We shall discuss in the next sections how drastically the failure criteria change the admissible scenarios of stable detachment velocity.

\subsection{Ideally brittle links}
\begin{figure}[ht] 
\captionsetup{justification=raggedright}
    \centering
    \includegraphics[width=.9\linewidth]{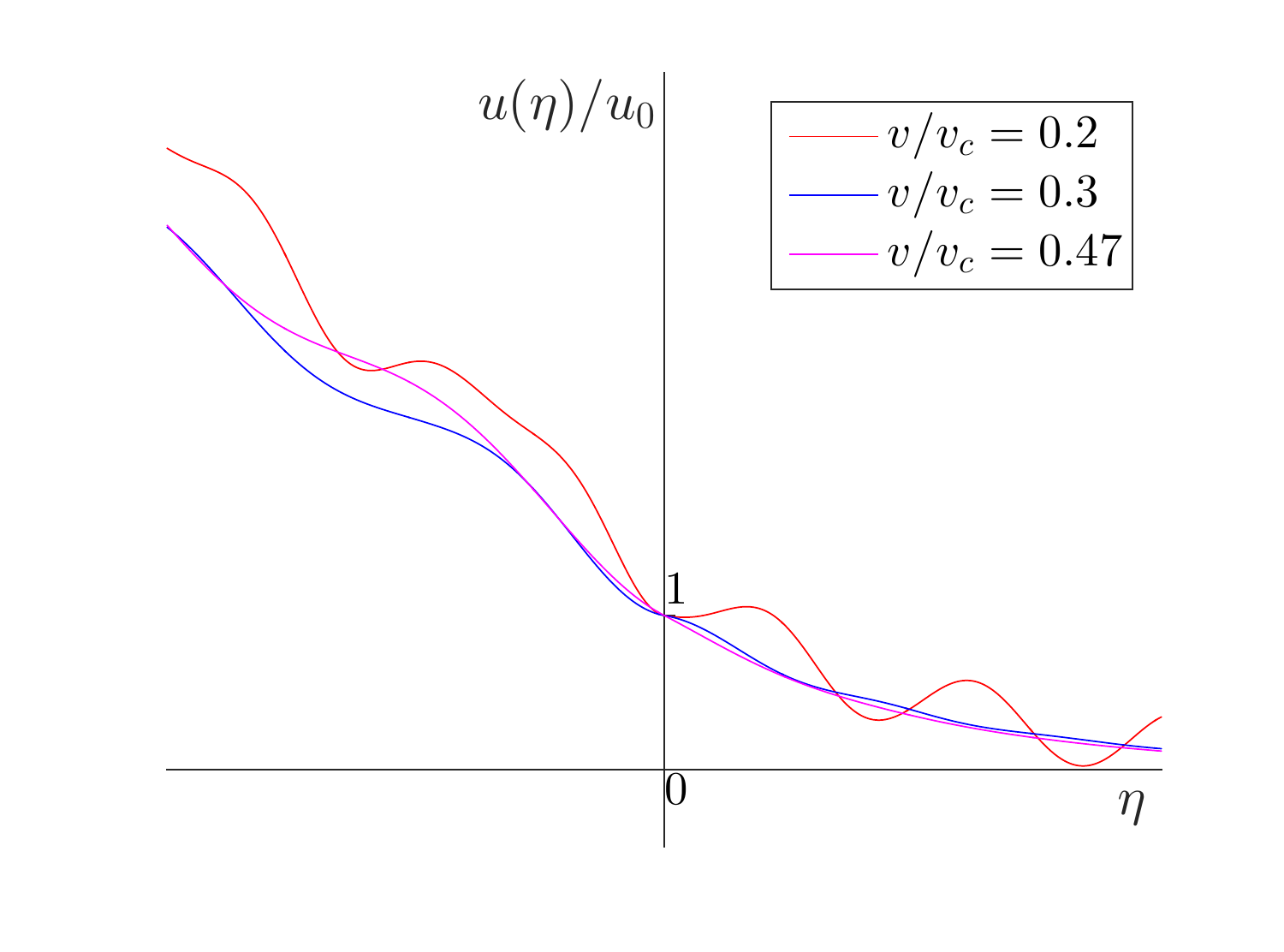} 
  \caption{Dimensionless steady-state displacements profiles. The crack speed $0.2 v_c$ shows $u(\eta)>u_s$ in the intact region $\eta>0$ ahead of the crack tip. It means that such velocity of the fracture wave is not admissible according to the instantaneous fracture criterion.}
  \label{fig:dispb}
\end{figure}

If the links are ideally brittle, the critical condition to be reached at the crack tip at the instant of fracture before further propagation
\begin{equation}
u_0=u_s
\label{eq:idbrit}
\end{equation}
is independent of the fracture propagation speed. With such a condition, that is when $u_0$ and $u_s$ are interchangeable, the diagrams of force and energy release rate in Figs.~\ref{fig:preq} are directly applicable. Furthermore, in this case, the condition of admissibility is easily checked i.e. that no points for $\eta>0$ are lifted higher than the crack tip. Looking at Fig.~\ref{fig:dispb}, one can notice that for such materials, configurations occurring at low $v$ are unphysical since there are points ahead of the crack tip where the failure criterion has been encountered already before the arrival of the fracture front itself and thus must be labelled as not admissible. Discussions on the matter have been dealt in \cite{marder1995origin,gorbushin2016analysis,gorbushin2017dynamic}. For example, the speed $0.2 v_c$ does not fulfill such requirement, then this must be discarded as non-admissible. On the contrary, the speeds $0.3 v_c$ and $0.47 v_c$ are admissible. Observing all the $u(\eta)$ profiles, for the isotropic chain the minimum velocity of the crack corresponds to about $0.27 v_c$ and all larger subsonic velocities are admissible (see Fig.~\ref{fig:admiss}). It is perhaps worth pointing out that such limit is smaller than the minimum energy release rate, which sits at about $0.38 v_c$. This implies that a single $G$ may correspond to two possible steady-states like $0.3 v_c$ and $0.47 v_c$. Such speeds, anyway, correspond to two different loads (see Fig.~\ref{fig:fvs1}). The highest of the two speeds is achieved uniquely by means of a larger force.

\section{Problem and methods}
When dealing with non-instantaneous criteria for fracture, the crack opening before fracture depends on the crack speed. In general,
\begin{equation}
u(0,v)=u_0(v)
\end{equation}
must be determined accordingly to the new fracture parameters and does not simply equal $u_s$ as for the ideal brittle criterion Eq.\eqref{eq:idbrit}. In this way, also the energy released locally at the crack tip $G_0(v)=k u^2_0(v)/2$ is a function of the crack speed. Further on we express $F$ as a multiple of the force required to break the spring in a static test via the function
\begin{equation}
\frac{F}{k u_s}=\Phi(v)\frac{u_s}{u_0(v)},
\label{eq:fove}
\end{equation}
which differs from the general Eq.\eqref{eq:fove0} where the denominator incorporates the elongation at failure, independently on the particular fracture criterion adopted. The reason is conceptual and follows from the possibility of conducting experiments. For obtaining the same crack speed, the loading condition, indeed, must be accurately designed depending on how the failure happens (i.e. within the context of this paper, which fracture criterion better describes the constituent material). We stick to the ratio $G/G_0$, instead of introducing a hypothetical $G_s=k u^2_s/2$, because the energy release rate incorporates the type of the structure and its deformation properties (in the present case a linear elastic chain) and can hardly be measured. {Moreover, in this way, as we will discuss further in the continuation, the dependency of $G/G_0$ remains untouched by the particular fracture criterion characterizing the links, while the force versus velocity relation depends on the particular criterion.}{}

\subsection{Incubation time criterion}
The incubation time failure criterion Eq.\eqref{eq:itor0} can be applied to the chain by a change of variable according to Eq.\eqref{eq:teta} which leads to
\begin{equation}
\Psi(v,\tau):=\frac{1}{v\tau}\dinteg{0}{v\tau}{\frac{u(\eta,v)}{u_0(v)}}{\eta}=\frac{u_s}{u_0}.
\label{eq:itchain}
\end{equation}
The normalization by the displacement at the crack tip $u_0$ is convenient because, in the ideally brittle case, the crack opening $u_0$ before fracture was known in every case and given by the maximum elongation criterion, now it is unknown and dependent on velocity. The shape $u(\eta,v)/u_0(v)$ of the deformation profile, though, is given once and for all as it does not depend on the particular value of the crack opening. The advantage is that, once one calculates the shape at a certain velocity from the solution of \cite{slepyan1984fracture}, this can be used for all the possible steady-state fracture criteria. If $\tau$ goes to zero, that is the material is ideally brittle, Eq.\eqref{eq:itchain} returns $u_0=u_s$ coherently. By this respect, one can say that IT materials are an extension of ideally brittle ones by means of $\tau$. Moreover, for steady-state propagation, the length $v\tau$ is constant in time and thus incubation time can be considered as a non-local criterion as well as a non-instantaneous one.

\subsection{Tuler-Butcher criterion}

In order to deal with the usual moving coordinate frame Eq.\eqref{eq:teta} and a steady-state regime of velocity $v$, the TB criterion Eq.\eqref{eq:TBcrit0} can be transformed into the equation
\begin{equation}
\dinteg{0}{+\infty}{\frac{\mathrm{H}(u(\eta,v)-u_s)}{vD}\frac{\left(u(\eta,v)-u_s\right)^2}{u_0(v)^2}}{\eta}=\frac{u_s^2}{u_0^2}.
\label{eq:tbcr}
\end{equation}
Taking advantage of the invariance of $u(\eta,v)/u_0(v)$ with respect to $\eta$, the function
\begin{equation}
\Lambda(v,D)=\frac{u_s}{u_0}
\label{eq:lambda}
\end{equation}
can be obtained as solution of Eq.\eqref{eq:tbcr}.

\section{Results}
In order to have dimensionless strength parameters, from this section on we express $u_s$ in units of the distance between the masses, whereas $D$ and $\tau$ are expressed in units of the same distance divided by the sound velocity $v_c$.
\begin{figure}[!ht]
\captionsetup{justification=raggedright}
\centering
\includegraphics[width=0.4\textwidth]{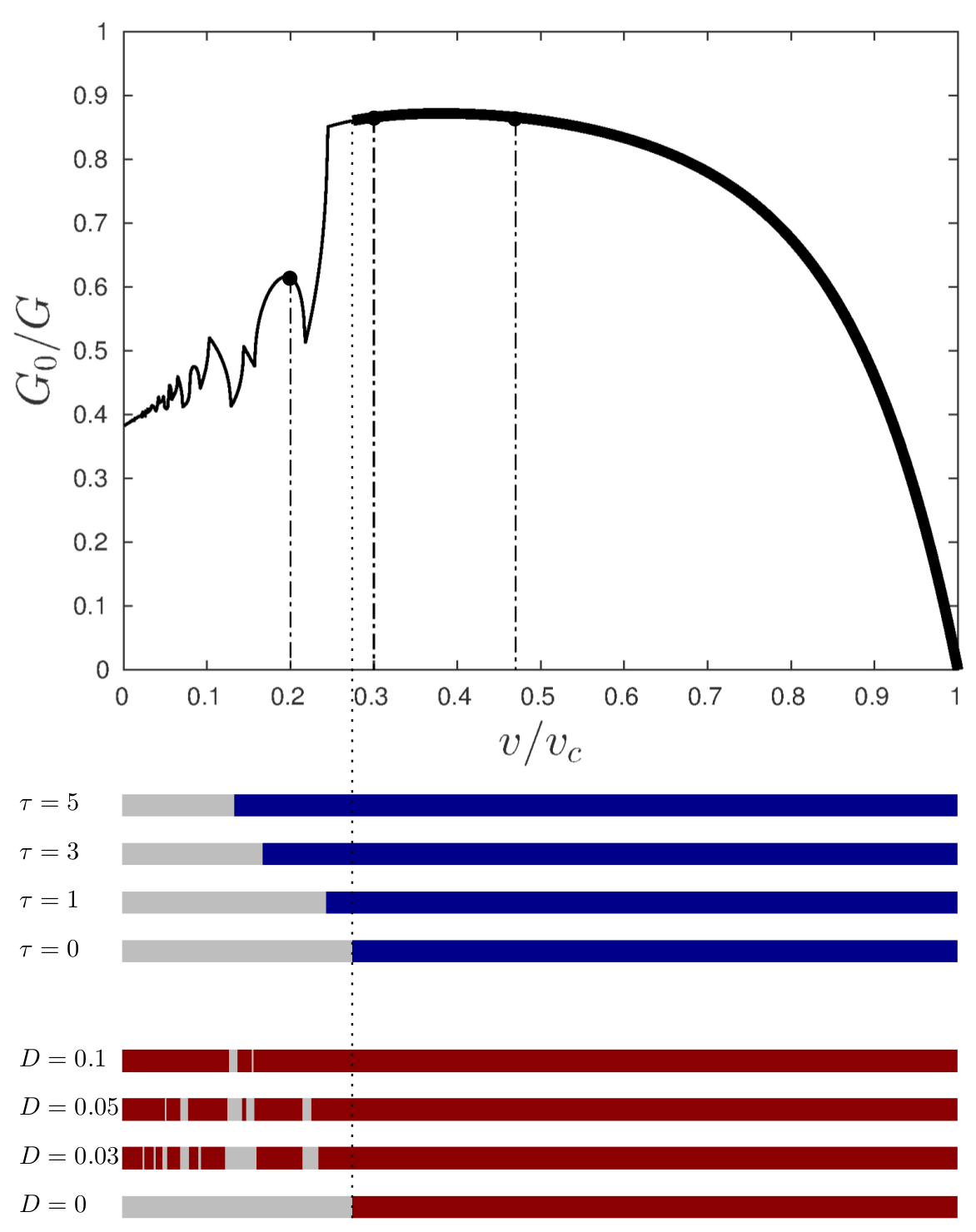}
\caption{Influence of fracture criteria on admissible regimes. The plot on the top shows the admissible regimes for ideally brittle links. Grey segments of the admissible bars indicate non-admissible crack speeds, blue bars refer to IT achievable steady-states, red bars to TB ones.}
\label{fig:admiss}
\end{figure}

\subsection{Incubation time criterion}
\begin{figure}[ht] 
\captionsetup{justification=raggedright}
  \begin{subfigure}[b]{.9\linewidth}
    \centering
    \includegraphics[width=\linewidth]{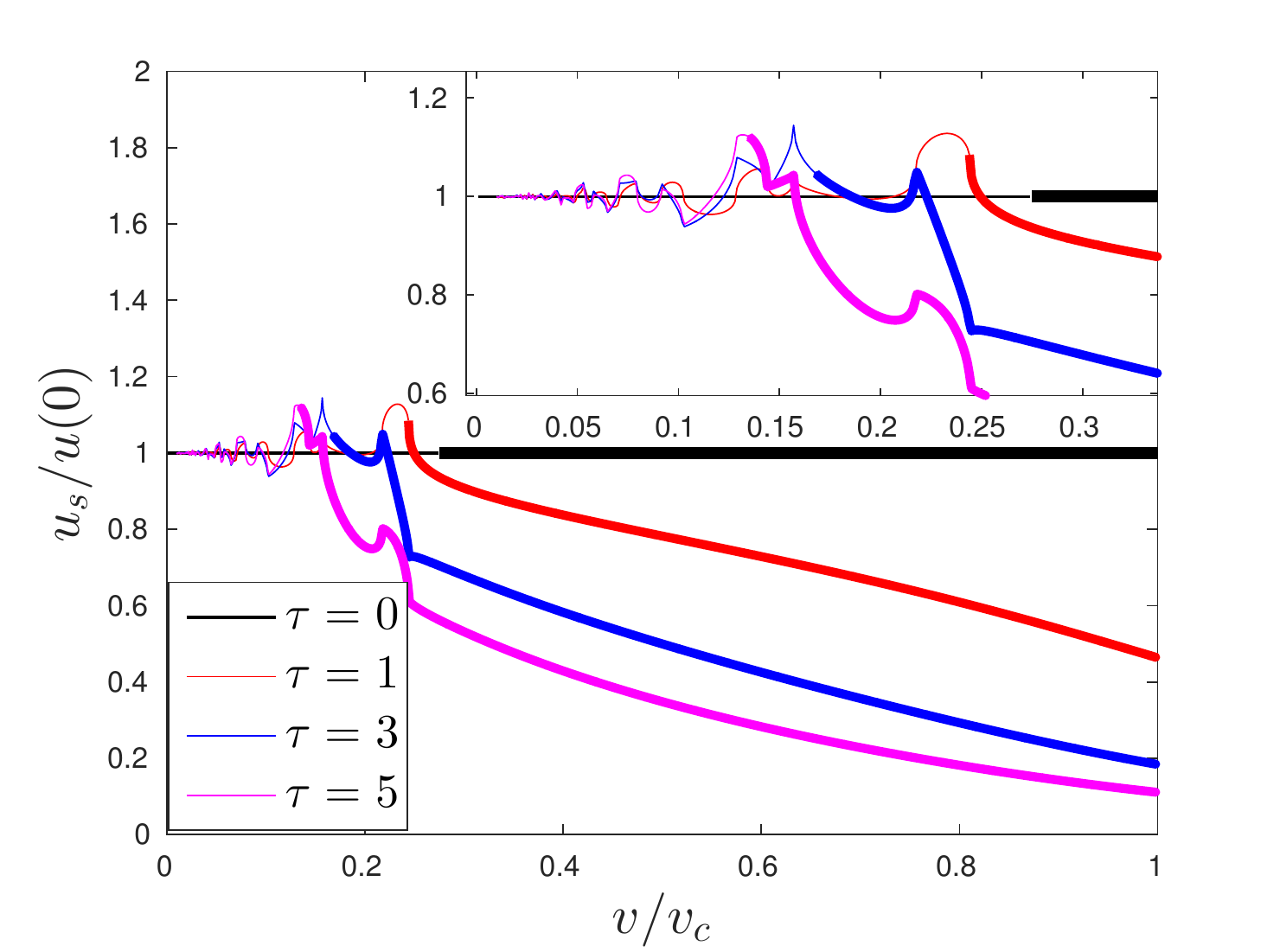} 
    \caption{Function $\Psi(v,\tau)$ versus velocity (see Eq.\eqref{eq:itchain}).}  
    \label{fig:usu0it} 
    \centering    
    \includegraphics[width=\linewidth]{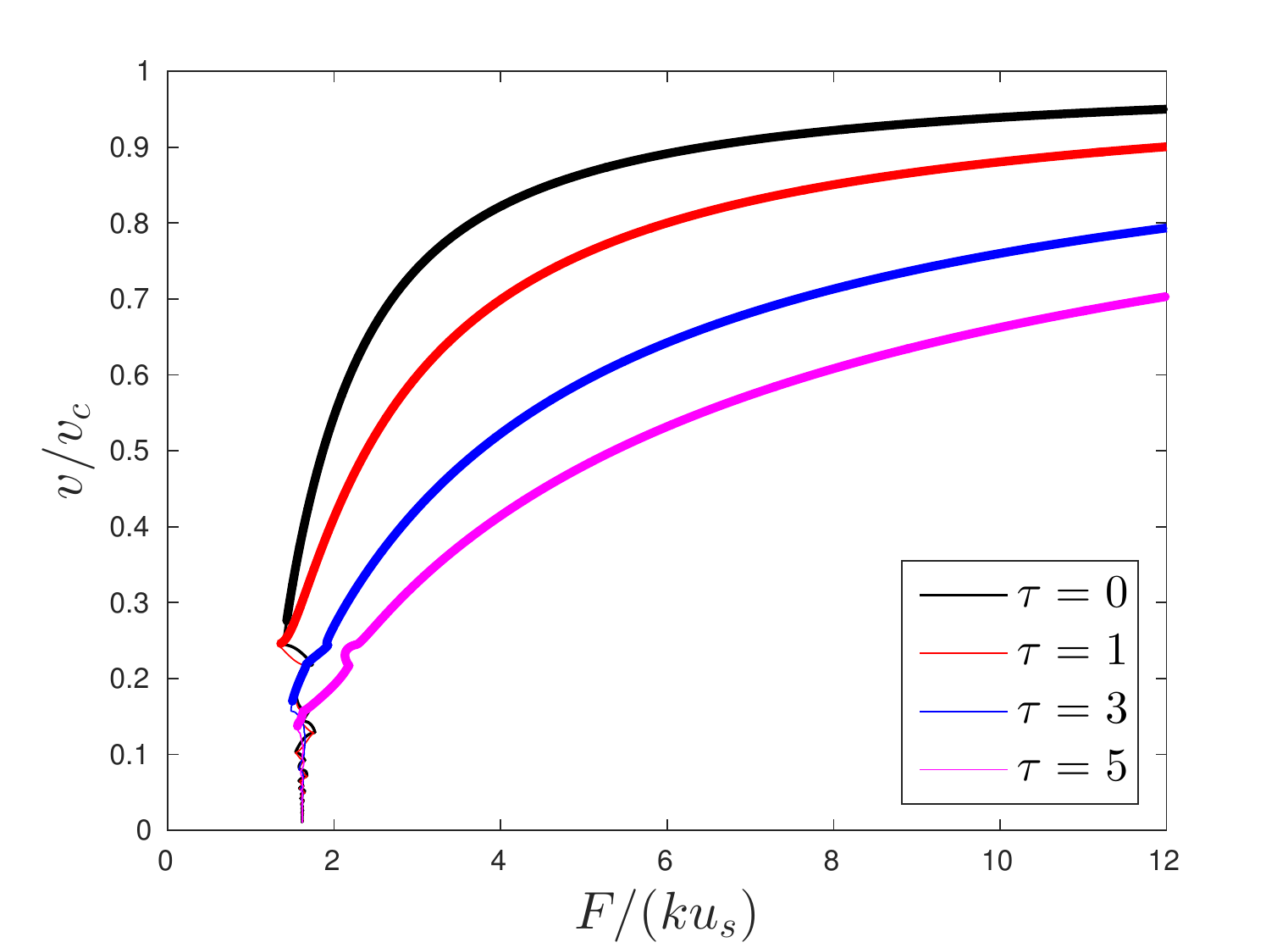}
    \caption{Force versus crack speed (see Eq.\eqref{eq:itforce}).} 
    \label{fig:fvsit} 
  \end{subfigure}
  \caption{Effects of incubation time $\tau$ on crack opening and applied force compared with the ideally brittle case. Thick lines stand for admissible regimes.  a) The function $\Psi(v,\tau)=u_s/u(0)=u_s/u_0$ measures how much the crack opening differs from the case of an ideally brittle chain. b) The crack speed dependent opening $u(0)=u(0,v)$ expressed through $\Psi(v,\tau)$ changes the prediction of the force to apply in order to cause a certain velocity.} 
\end{figure}

The steady-state analogue of the incubation time criterion in Eq.\eqref{eq:itchain}, which also defines the function $\Psi(v,\tau)$, solves the issue of calculating the crack opening, given $\tau$ and $u_s$
\begin{equation}
u_0=\frac{u_s}{\Psi(v,\tau)}.
\label{eq:uous}
\end{equation}
Speaking of the force to apply for achieving a certain steady-state velocity, substituting Eq.\eqref{eq:uous} into Eq.\eqref{eq:fove} gives
\begin{equation}
\frac{F}{k u_s}=\frac{\Phi(v)}{{\Psi(v,\tau)}}.
\label{eq:itforce}
\end{equation}

\begin{figure}[ht] 
\captionsetup{justification=raggedright}
    \centering
    \includegraphics[width=0.4\textwidth]{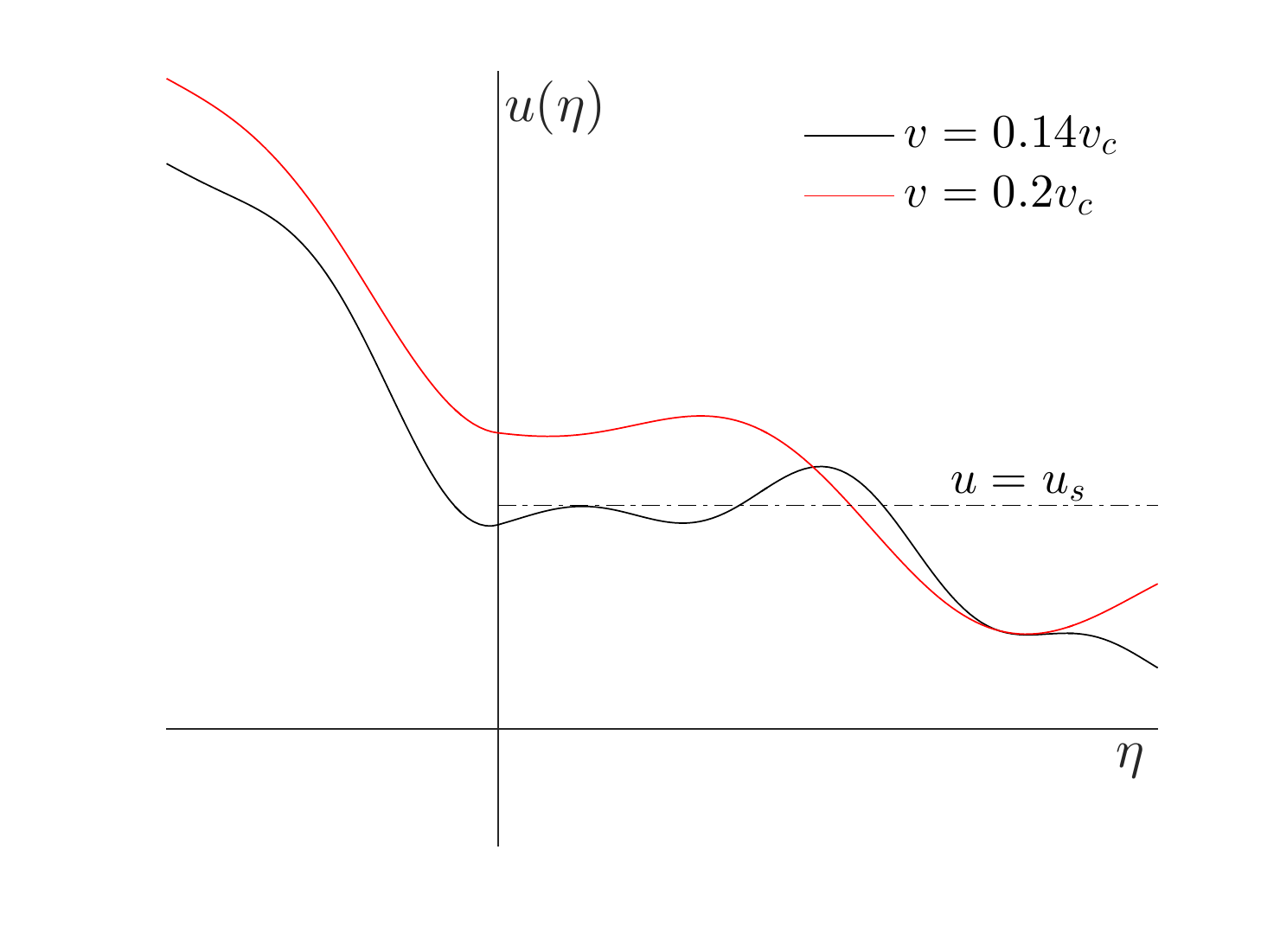}        
  \caption{Crack opening for small velocities and high incubation times. With $\tau=5$, the crack tip opening $u_0$ for the admissible speed $v=0.14 v_c$ is predicted to be smaller than the static strength $u_s$. For comparison, $v=0.2 v_c$ shows the most common situation of $u_0>u_s$. See also Fig.~\ref{fig:usu0it}.} 
  \label{fig:itdipara}
\end{figure}

The behavior of the function $\Psi(v,\tau)$ influences the way $\tau$ modifies the crack opening with respect to the ideally brittle one ($\tau\rightarrow 0$) and this is shown in Fig.\ref{fig:usu0it}. A linear elastic bond which exhibits a non-zero incubation time will in general allow a bigger crack opening at the instant of fracture. Crack openings smaller than $u_s$, though, are admissible at low velocities due to rapid oscillations and negative $\partial u/\partial t$ close to the tip (see Fig.\ref{fig:itdipara}).
The influence of $\tau$ on the force is plotted in Fig.\ref{fig:fvsit}. Given the result of the static test on the spring $u_s$, if the goal is achieving a certain velocity $v$, an IT type material predicts that the steady-state regime would be reached in general via a bigger force than one could expect if $\tau$ is neglected. The region where the relation between force and velocity is not bijective is stretched to the right and the difference in velocities for the same force decreases steadily while raising the incubation time. Note that $\tau$ does not play any role in the limiting case of zero velocity of propagation, where the force  produces the same results as for the ideally brittle spring (see Eq.\eqref{eq:itchain} for $v\rightarrow 0$).
In order to verify the analytical solution, we searched for the steady-state regimes via solving Eq.\eqref{eq:orig} by a finite difference scheme similarly as in \cite{gorbushin2017dynamic} where the same numerical procedure is extensively explained. A chain of 2000 masses was loaded with a distant vertical constant force. The instant of fracture for a link was identified according to the condition Eq.\eqref{eq:itor0}. After iterating, the next failures tended to occur at constant time pace and such an interval was used for calculating the stable crack speed for a given force. The analytical solution is perfectly matched and the numerical approach confirms that a steady-state propagation is not achievable at low velocities (results not shown).

This kind of computation is quite heavy because of the algorithm adopted to identify the time of fracture at every location $x_i$. Before a steady fracture propagation is reached, indeed, during the transient regime, the history $u_i(t)$ in the last interval $\tau$ must be recorded and the integral Eq.\eqref{eq:TBcrit0} updated for every $x_i$. This marks a principal difference with the instantaneous traditional criterion in which case the quick check $u\leq u_s$ is sufficient. We have tried to simplify the procedure via making the criterion pseudo-instantaneous. Theoretically, Eq.\eqref{eq:uous} should be valid only for steady-state fracture. Nevertheless, if one estimates the instantaneous crack velocity $\tilde{v}(t)$ from the last two failures, the criterion can be artificially reduced to $u\leq u_0(\tilde{v},\tau)$ instead of calculating the integral Eq.\eqref{eq:uous} at all. Such an attempt has been proven to be effective beyond expectations for the particular studied problem in achieving the same steady-states as in theory and the rigorous numerical simulation for same applied load.

Another crucial effect of $\tau>0$ on the crack propagation is that it monotonically enlarges the regions of achievable steady-states as illustrated in Fig.~\ref{fig:admiss}. For instance, the speed $0.2 v_c$, that is non-admissible for an ideally brittle material with critical elongation parameter $u_s$, can be reached with $\tau=3$ and bigger.

\subsection{Tuler-Butcher criterion}
\begin{figure}[ht] 
\captionsetup{justification=raggedright}
  \begin{subfigure}[b]{.9\linewidth}
    \centering
    \includegraphics[width=\linewidth]{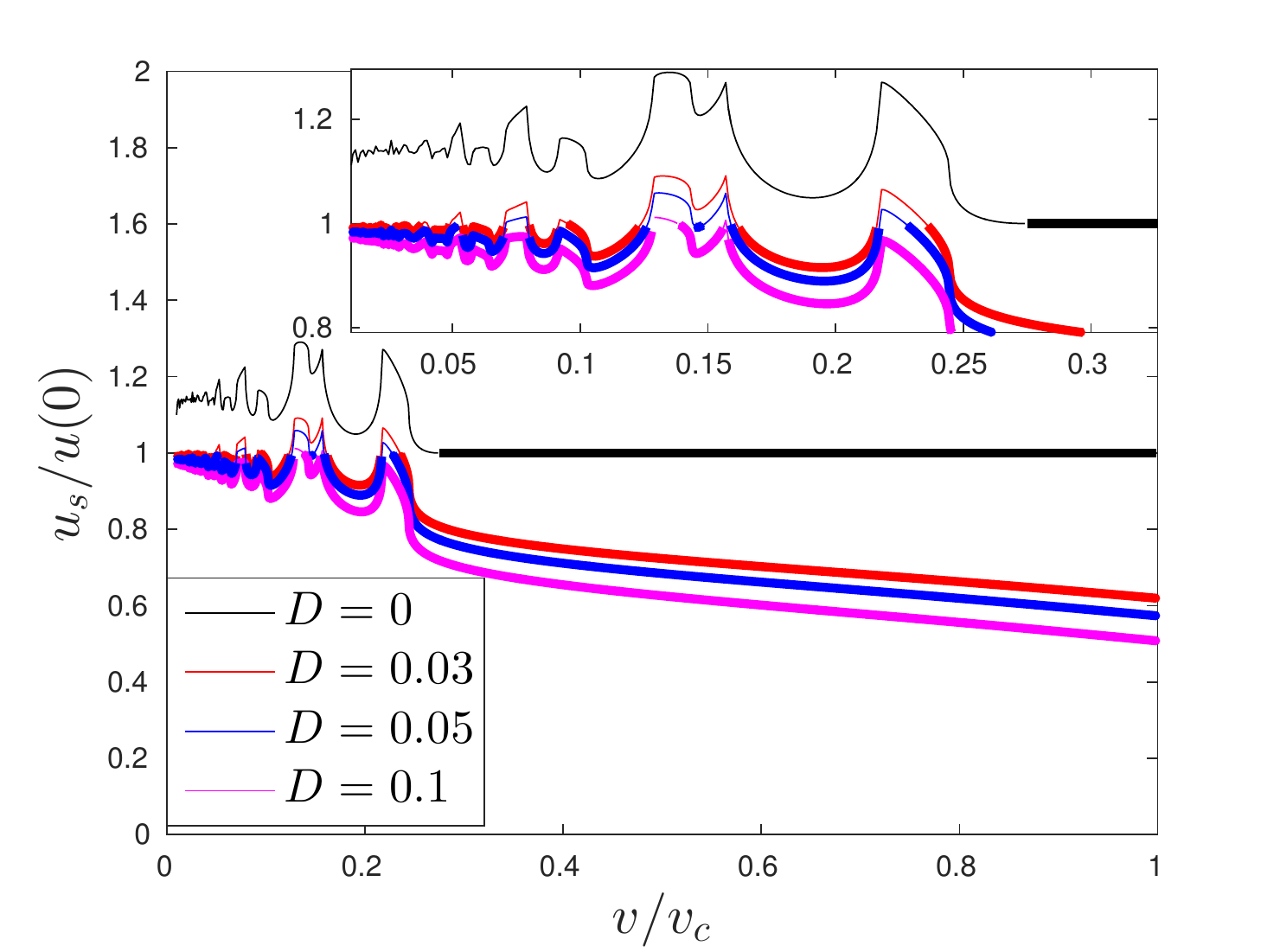} 
    \caption{Function $\Lambda(v,D)$ versus velocity (see Eq.\eqref{eq:tbcr}).}  
    \label{fig:usu0tb} 
    \centering    
    \includegraphics[width=\linewidth]{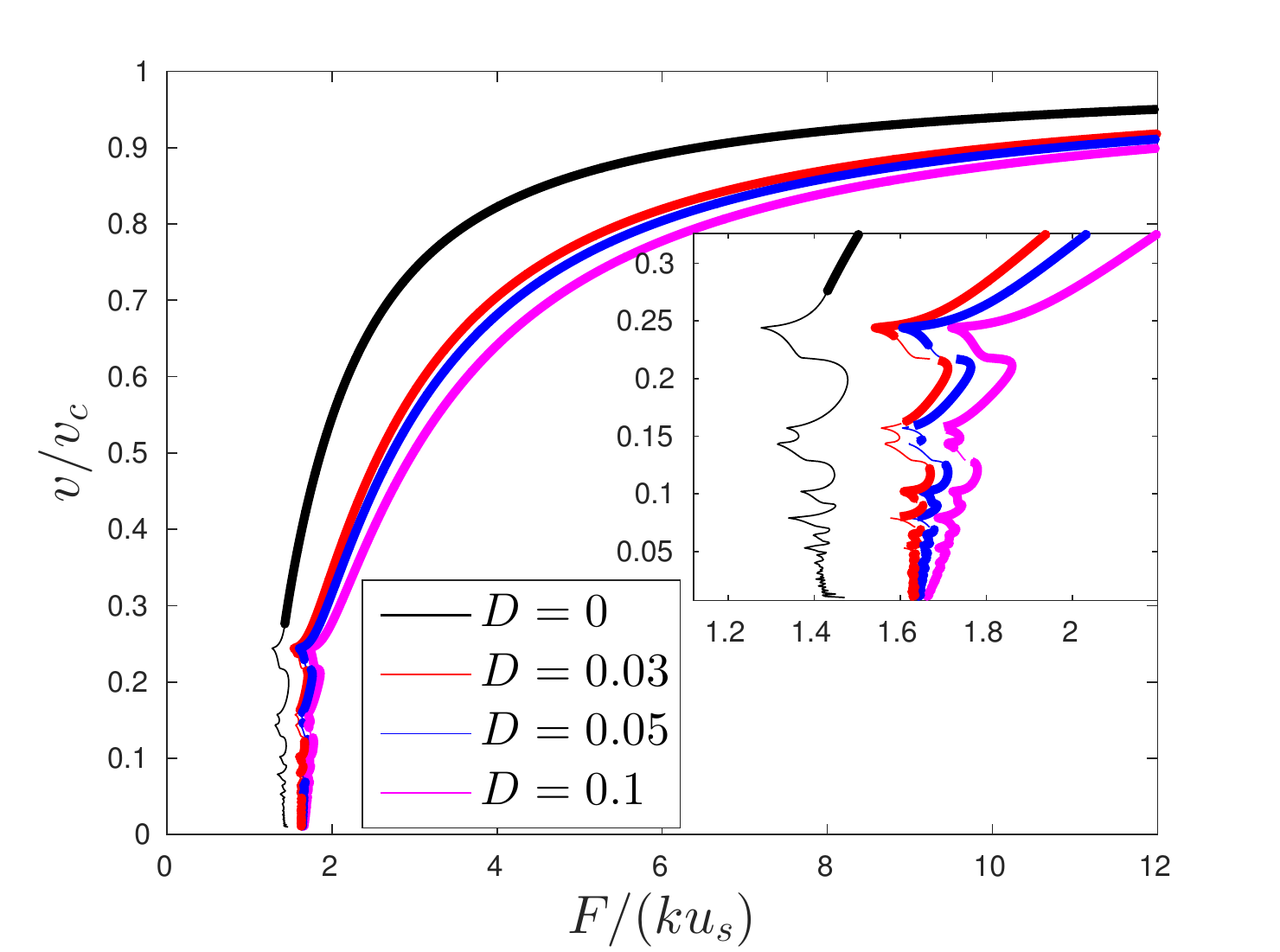}
    \caption{Force versus crack speed (see Eq.\eqref{eq:TBforce}).} 
    \label{fig:fvstb} 
  \end{subfigure}
  \caption{Effects of overload maximum work $D$ on crack opening and applied force compared with the ideally brittle case. Thin lines represent unphysical steady-states. a) The function $u_s/\Lambda(v,D)$ measures how much the crack opening differs from an ideally brittle chain. b) The crack speed dependent opening $u_0$ expressed through $\Lambda(v,D)$ changes the prediction of the force to apply in order to cause a certain velocity.} 
  \label{fig:uFtb}
\end{figure}
By means of Eq.\eqref{eq:lambda}, one can retrieve the crack opening
\begin{equation}
u_0=\frac{u_s}{\Lambda(v,D)}
\end{equation}
associated to all the combinations of crack speed and $D$. As consequence, keeping the force proportional to the material property $u_s$ instead of the velocity dependent crack opening, Eq.\eqref{eq:fove} becomes
\begin{equation}
\frac{F}{k u_s}=\frac{\Phi(v)}{\Lambda(v,D)}.
\label{eq:TBforce}
\end{equation}

\begin{figure}[ht] 
\captionsetup{justification=raggedright}
    \centering
    \includegraphics[width=0.4\textwidth]{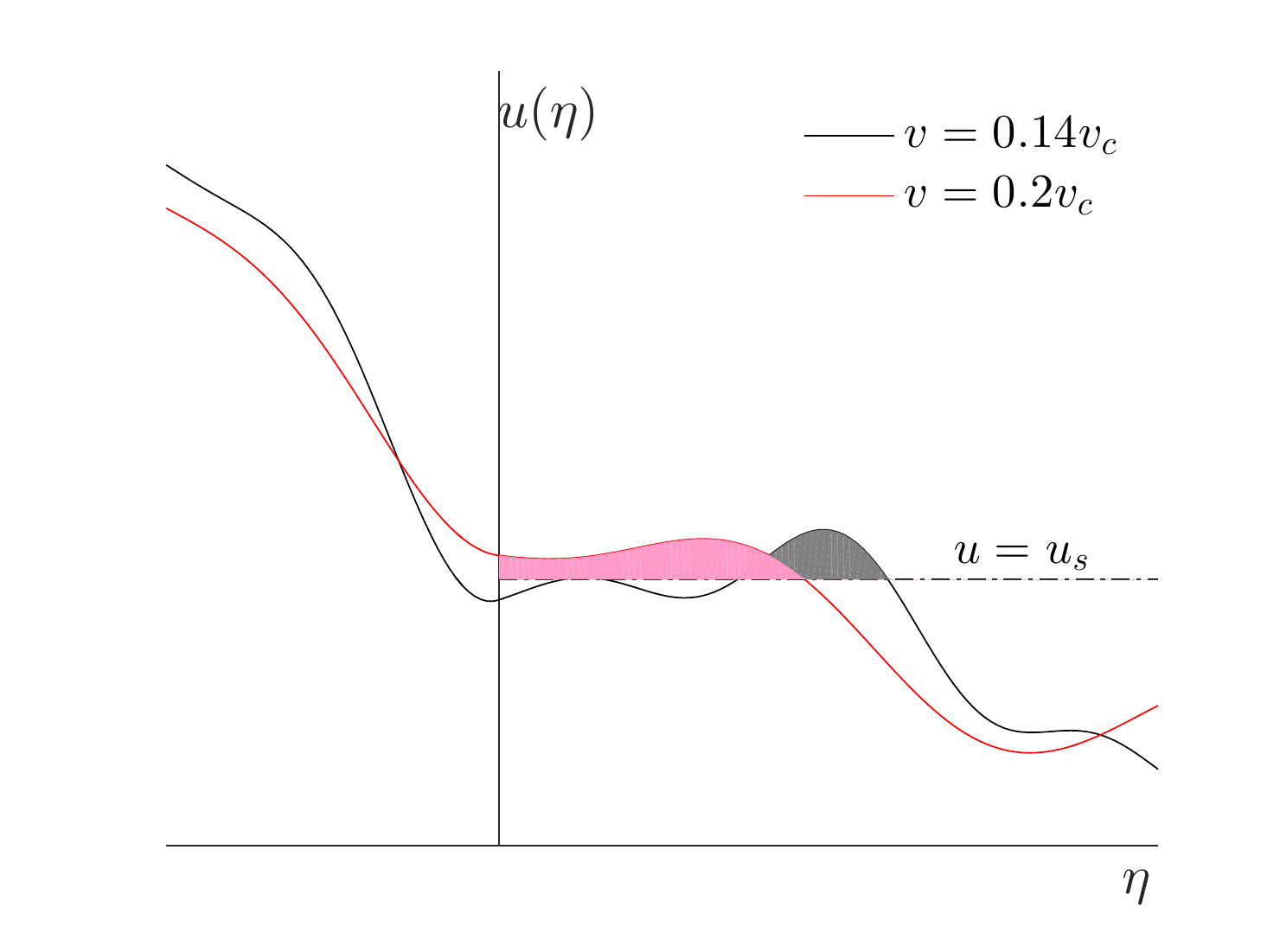}        
  \caption{Crack opening at low velocities for small $D$. The profile $u(\eta)$ for two fracture speeds at $D=0.03$ is shown. The shaded faces represent the areas where the integral Eq.\eqref{eq:tbcr} must be computed. The result for $v=0.14v_c$ shows how $u_0<u_s$ means that the chain detaches ahead of $\eta=0$ making that crack speed non-admissible. See also Fig.\ref{fig:usu0tb}.} 
  \label{fig:tbdipara}
\end{figure}

The plots in Fig.\ref{fig:uFtb} allow us to visualize how the dynamic strength parameter $D$ affects the chain behavior. As observed for IT materials, the immediate impact of TB damage accumulation results in an augmented crack opening at equal crack speed as an ideally brittle material showing the same static strength $u_s$, at least in the range of medium or high $v/v_c$ (see Fig.\ref{fig:usu0tb}). The force needed for obtaining a desired velocity is depicted in Fig.\ref{fig:fvstb}. It is evident that the capability of the material to bear a certain work of the overstretch before failure, i.e. bigger $D$, makes the chain detachment increasingly slower for same $F/k u_s$. A structure of TB bonds can be predicted to be dynamically tougher than its ideally brittle counterpart. For $D\rightarrow 0$ and low fracture speed the complex structure does not respond like an ideally brittle one. Looking at Fig.\ref{fig:tbdipara}, indeed, one can notice that for low $v$, i.e. in the range where $u(\eta)$ does not decrease monotonically ahead of the crack tip, the criterion in the form Eq.\eqref{eq:tbcr} may return $u_0<u_s$. Such a feat differs from all the cases analyzed in this work: single bonds as well as the ideally brittle and IT crack tips always fail for $u\geq u_s$. However, in the cases where the latter inequality is violated in a TB link, we obtain non-admissible steady-state propagation regimes. A point can be made, therefore, that a theoretical limit for the crack opening is for a TB material to satisfy $u_0\geq u_s$.

Like for incubation time materials, as it can be seen in Fig.~\ref{fig:admiss}, new zones of admissibility appear in the low velocity region for larger $D$. Nevertheless, there is a significant qualitative difference with the incubation time situation: such admissible intervals emerge small and scattered, but then, with increasing $D$, expand gradually and merge until every subsonic crack speed can be obtained for $D$ close to unity.

\section{Discussion and conclusions}
The dynamic fracture propagation in discrete structures has been investigated in a considerable amount of possible scenarios (see references above) but the influence of failure criteria different from a threshold stress has not gained the attention that it deserves despite the fact that non-instantaneous criteria have already been shown to be reliable in continuum mechanics (e.g. as recently discussed in \cite{Alves2017}). As a first step to fill this gap, two time-dependent criteria have been analyzed in detail when applied to the dynamic fracture propagation of a chain of oscillators and they have been compared to the classical ideal brittle fracture.
In both cases, enhanced admissibility have appeared at low crack speeds and mapped in Fig.~\ref{fig:admiss}. An increasing incubation time $\tau$ enlarges the admissibility continuously but never covers all the subsonic crack speeds. More than that, a TB material which requires a larger work of the overstretch for fracture and characterized by a bigger $D$ also creates completely new zones of achievable steady-states and it is predicted that all the subsonic range is possible if $D\gtrsim 0.13$.

Speaking of the steady-state crack opening, the time-dependent criteria cause a delay in fracture after reaching the static strength of the bonds. This means that in most cases one should expect $u_0>u_s$ like it would happen when monotonically elongating a single spring. At low $v$, though, this is not the behavior caused by a constant force applied on a complex structure. Ample and rapid oscillations ahead of the crack tip cause the delayed fracture to happen at $u_0<u_s$. While such propagation regimes are admissible at high $\tau$ for incubation materials, the same is not true for TB ones (see Figs.\ref{fig:itdipara}-\ref{fig:tbdipara}). The mathematical form of the latter failure criterion indeed excludes such steady-states on the grounds that daughter cracks would jeopardize the steady-state assumption. In short, a theoretical limit has been found which states that for a TB chain a dynamic fracture can propagate at constant speed only if $u_0\geq u_s$.

As intuition suggests, the two examined non-instan\-taneous criteria make the structure tougher than the corresponding ideal brittle one with the same static strength $u_s$. Thus, for obtaining a certain velocity $v$ one needs a bigger force if $\tau$ or $D$ increase. In this way, the curves in Figs.\ref{fig:fvsit}-\ref{fig:fvstb} result also in an important practical application. With a few experiments on materials whose $u_s$ has been independently measured, the couples $F$-$v$ allow for the material characterization in terms of the second fracture parameter $\tau$ or $D$ at least if stable crack speeds are retrieved in the monotonic interval of the curves (medium and high $v/v_c$).

The velocity dependent energy release rate ratio $G/G_0$ is a solution which is irrespective of the particular fracture criterion adopted. It is also valid regardless of the way the energy is introduced into the system. In the present work we use a constant force as the external load. However, in case one prefers to implement different kind of loading, like for instance in \cite{marder1995origin} when dealing with lattices, or for example to facilitate a specific experimental procedure, the relation between the new load and the crack speed has to be evaluated, while the energy - crack speed diagram remains the same. 

Two numerical integration schemes have been used to solve the set of governing equations in the IT scenario and compared with the analytical solution derived in the present paper. The first of these verified the criterion condition in its integral form Eq.\eqref{eq:itor0} at every time step for all the unbroken bonds. Such an approach enables one to simulate also the transient regimes before a steady-state is reached. The steady-states were achieved only in the admissible regions of Fig.\ref{fig:admiss} and there the force-velocity relations agreed perfectly with the ones in Fig.\ref{fig:fvsit}. The second test was performed by adopting a pseudo-static failure criterion: a dynamic threshold elongation $u_0(\tilde{v},\tau)$ was established based on the instantaneous crack velocity $\tilde{v}$ and Eq.\eqref{eq:uous}. This simplified algorithm performed much faster than the first one, and still returned correct results. For this specific structure at least then, it seems that many of the conclusions drawn here can still be valid in the transient regimes.

{Beyond the particular scope of this paper, various propagation regimes, in absence of crack arrest, can appear: steady-state, other regular ones (clustering or forerunning as discussed in \cite{mishuris2008dynamics,mishuris2009localised,slepyan2010crack,slepyan2015forerunning,nieves2016analysis,nieves2017transient}) or chaotic regimes. The realization of one or another heavily depends on the loading type, its intensity and on the structure itself. However, when the problem is faced from a mathematical point of view, assumptions of steady-state regimes have always been made in order to obtain simple solutions. With the analytical results in one hand, an a posteriori examination is required which identifies where the solution fulfills the assumptions and constraints: only that part of the solution is labeled as \textit{admissible}. Generally speaking, though, it does not mean that all those regimes will emerge in practice as steady-state. In most of the cases it happens; nevertheless, as it has been shown in numerical simulations, other ordered regimes of propagation may arise such as clustering. In such circumstances, the steady-state velocity predicted theoretically reveals as the average speed at which the cluster moves (see \cite{mishuris2009localised,nieves2016analysis,nieves2017transient} on the matter).}{}

We have not treated the problem of branching in the present settings of history-dependent criteria. It has turned out already in \cite{marder1995origin,gorbushin2016cer,gorbushin2017dynamic} that such instabilities can become relevant at high crack speeds. In the considered geometry, loading condition and material parameters, the branching mostly happens 
along the crack surfaces but not on the crack line ahead. That is evident from the solution profile $u(\eta)$ for $v/v_c=0.2$ in Fig.\ref{fig:dispb}. If the horizontal springs show the same dynamic resistance as the vertical ones, their fracture can precede the chain detachment from the substrate, making a steady-state propagation impossible {and sensibly reducing the limiting speed with respect to $v_c$}{}. The admissibility check would imply that, for none of the consecutive oscillators, the difference $u_{i+1}-u_i)$ does reach the condition imposed by the fracture criterion. More complex scenarios may occur with structures characterized by flexural stiffness, heterogeneities or localized feeding waves \cite{mishuris2009localised,nieves2016analysis,nieves2017transient}. Furthermore, a steady-state regime can be unattainable, resulting in unstable or alternating velocities, when the structure is not loaded far from the crack tip, but via accumulated energy in the form of residual stresses of the bonds \cite{ayzenberg2014brittle}. Complications have also been object of investigation in the framework of bridged cracks \cite{mishuris2008dynamics}.

In conclusion, the fracture criteria considered here sensibly affect the dynamic propagation of cracks in discrete structures. The effects are particularly important both in terms of force vs velocity relations and in new regimes of admissibility at low crack speeds. Succinctly, by the present results, we are able to underline two main messages:
\begin{itemize}
\item[\textbullet] it is at low speed regimes that an experimental investigation should be carried out more carefully for understanding whether it is necessary to incorporate history-dependent fracture criteria in the dynamic fracture model;
\item[\textbullet] the energy release rate ratio and shapes of the displacement profiles as functions of the velocity are invariants, in linear theory, and can promptly be used and adapted to the most suitable fracture criterion for the analyzed problem.
\end{itemize}

The possible outlook of this research is the application of the approach to a) more complex lattice structures (inhomogeneous, triangle ones) as in \cite{Mishuris_Slepyan2007,Nieves_Mish2013};
b) highly ordered bi-dimensional lattices, for instance to crack propagation in graphene layers \cite{xiao2009fracture,tsai2010characterizing} or c) to the unbinding of long protein chains whose analysis has been made feasible by the improvements in the field of atomic force microscopy and for which the bonds strength has already shown to be eminently dependent on the strain rate \cite{merkel1999energy,hinterdorfer2006detection}.
As the discretization, the considered here model can be useful for modeling of the peel test of flexible films (see e.g. \cite{Wei1998,Nase_Eremeev2016}).

\section*{Acknowledgements}
N. Gorbushin and G. Vitucci acknowledge support from the EU project CERMAT2 (PITN-GA-
2013-606878), G. Mishuris and G. Volkov are thankful to the EU project TAMER (IRSES-GA-2013-610547). G. Mishuris also thanks the UK Royal Society Wolfson Research Merit Award, while G. Volkov received support also from RFBR (17-01-00618, 16-51-53077). {We wish to express our gratitude to Michael Nieves for interesting discussions.}{}

\bibliographystyle{ieeetr}
\bibliography{Bibliography}

\begin{thebibliography}{10}

\bibitem{abraham1998spanning}
F.~F. Abraham, J.~Q. Broughton, N.~Bernstein, and E.~Kaxiras, ``Spanning the
  continuum to quantum length scales in a dynamic simulation of brittle
  fracture,'' {\em EPL (Europhysics Letters)}, vol.~44, no.~6, p.~783, 1998.

\bibitem{atrash2011evaluation}
F.~Atrash and D.~Sherman, ``Evaluation of the thermal phonon emission in
  dynamic fracture of brittle crystals,'' {\em Physical Review B}, vol.~84,
  no.~22, p.~224307, 2011.

\bibitem{bolander2005irregular}
J.~Bolander and N.~Sukumar, ``Irregular lattice model for quasistatic crack
  propagation,'' {\em Physical Review B}, vol.~71, no.~9, p.~094106, 2005.

\bibitem{tsai2010characterizing}
J.-L. Tsai, S.-H. Tzeng, and Y.-J. Tzou, ``Characterizing the fracture
  parameters of a graphene sheet using atomistic simulation and continuum
  mechanics,'' {\em International Journal of Solids and Structures}, vol.~47,
  no.~3, pp.~503--509, 2010.

\bibitem{glassmaker2007biologically}
N.~J. Glassmaker, A.~Jagota, C.-Y. Hui, W.~L. Noderer, and M.~K. Chaudhury,
  ``Biologically inspired crack trapping for enhanced adhesion,'' {\em
  Proceedings of the National Academy of Sciences}, vol.~104, no.~26,
  pp.~10786--10791, 2007.

\bibitem{marder1995origin}
M.~Marder and S.~Gross, ``Origin of crack tip instabilities,'' {\em Journal of
  the Mechanics and Physics of Solids}, vol.~43, no.~1, pp.~1--48, 1995.

\bibitem{bernstein2003lattice}
N.~Bernstein and D.~Hess, ``Lattice trapping barriers to brittle fracture,''
  {\em Physical review letters}, vol.~91, no.~2, p.~025501, 2003.

\bibitem{kessler2003does}
D.~A. Kessler and H.~Levine, ``Does the continuum theory of dynamic fracture
  work?,'' {\em Physical Review E}, vol.~68, no.~3, p.~036118, 2003.

\bibitem{Bitzek2015}
E.~Bitzek, J.~R. Kermode, and P.~Gumbsch, ``Atomistic aspects of fracture,''
  {\em International Journal of Fracture}, vol.~191, no.~1, pp.~13--30, 2015.

\bibitem{Morini2013}
L.~Morini, A.~Piccolroaz, G.~Mishuris, and E.~Radi, ``On fracture criteria for
  dynamic crack propagation in elastic materials with couple stresses,'' {\em
  International Journal of Engineering Science}, vol.~71, pp.~45--61, 2013.

\bibitem{Mishuris_Picci2012}
G.~Mishuris, E.~Radi, and A.~Piccolroaz, ``Steady-state propagation of a mode
  iii crack in couple stress elastic materials,'' {\em International Journal of
  Engineering Science}, vol.~61, pp.~112--128, 2012.

\bibitem{gorbushin2016analysis}
N.~Gorbushin and G.~Mishuris, ``Analysis of dynamic failure of the discrete
  chain structure with non-local interactions,'' {\em Mathematical Methods in
  the Applied Sciences}, 2016.

\bibitem{ravi1984experimental}
K.~Ravi-Chandar and W.~Knauss, ``An experimental investigation into dynamic
  fracture: {III}. {O}n steady-state crack propagation and crack branching,''
  {\em International Journal of Fracture}, vol.~26, no.~2, pp.~141--154, 1984.

\bibitem{fineberg1991instability}
J.~Fineberg, S.~P. Gross, M.~Marder, and H.~L. Swinney, ``Instability in
  dynamic fracture,'' {\em Physical Review Letters}, vol.~67, no.~4, p.~457,
  1991.

\bibitem{hauch1999dynamic}
J.~A. Hauch, D.~Holland, M.~Marder, and H.~L. Swinney, ``Dynamic fracture in
  single crystal silicon,'' {\em Physical Review Letters}, vol.~82, no.~19,
  p.~3823, 1999.

\bibitem{brun2014transition}
M.~Brun, G.~F. Giaccu, A.~B. Movchan, and L.~I. Slepyan, ``Transition wave in
  the collapse of the san saba bridge,'' {\em Frontiers in Materials}, vol.~1,
  p.~12, 2014.

\bibitem{stick09}
K.~King, ``The kinetic king detonates a guinness world-record stick bomb --
  2250 sticks!.''

\bibitem{Petrov_Utkin89}
Y.~V. Petrov and A.~Utkin, ``Dependence of the dynamic strength on loading
  rate,'' {\em Materials Science}, vol.~25, no.~2, pp.~153--156, 1989.

\bibitem{volkov2017some}
G.~Volkov, Y.~V. Petrov, and A.~Utkin, ``On some principal features of data
  processing of spall fracture tests,'' {\em Physics of the Solid State},
  vol.~59, no.~2, pp.~310--315, 2017.

\bibitem{Petrov2004}
Y.~V. Petrov, ``Incubation time criterion and the pulsed strength of continua:
  fracture, cavitation, and electrical breakdown,'' in {\em Doklady Physics},
  vol.~49, pp.~246--249, Springer, 2004.

\bibitem{Gruzdkov_Petrov}
A.~Gruzdkov and Y.~V. Petrov, ``Cavitation breakup of low-and high-viscosity
  liquids,'' {\em Technical Physics}, vol.~53, no.~3, pp.~291--295, 2008.

\bibitem{tuler1968criterion}
F.~R. Tuler and B.~M. Butcher, ``A criterion for the time dependence of dynamic
  fracture,'' {\em International Journal of Fracture Mechanics}, vol.~4, no.~4,
  pp.~431--437, 1968.

\bibitem{boustie1991experimental}
M.~Boustie and F.~Cottet, ``Experimental and numerical study of laser induced
  spallation into aluminum and copper targets,'' {\em Journal of applied
  physics}, vol.~69, no.~11, pp.~7533--7538, 1991.

\bibitem{wei2009modeling}
S.~Wei, F.~Qun-bo, W.~Fu-chi, and M.~Zhuang, ``Modeling of micro-crack growth
  during thermal shock based on microstructural images of thermal barrier
  coatings,'' {\em Computational Materials Science}, vol.~46, no.~3,
  pp.~600--602, 2009.

\bibitem{grady2010length}
D.~E. Grady, ``Length scales and size distributions in dynamic fragmentation,''
  {\em International Journal of Fracture}, vol.~163, no.~1, pp.~85--99, 2010.

\bibitem{slepyan1984fracture}
L.~Slepyan and L.~Troyankina, ``Fracture wave in a chain structure,'' {\em
  Journal of Applied Mechanics and Technical Physics}, vol.~25, no.~6,
  pp.~921--927, 1984.

\bibitem{slepyan2012models}
L.~I. Slepyan, {\em Models and phenomena in fracture mechanics}.
\newblock Springer Science \& Business Media, 2012.

\bibitem{slepyan1999lattice}
L.~I. Slepyan, M.~Ayzenberg-Stepanenko, and J.~P. Dempsey, ``A lattice model
  for viscoelastic fracture,'' {\em Mechanics of Time-Dependent Materials},
  vol.~3, no.~2, pp.~159--203, 1999.

\bibitem{kresse2004lattice}
O.~Kresse and L.~Truskinovsky, ``Lattice friction for crystalline defects: from
  dislocations to cracks,'' {\em Journal of the Mechanics and Physics of
  Solids}, vol.~52, no.~11, pp.~2521--2543, 2004.

\bibitem{nieves2016analysis}
M.~Nieves, G.~Mishuris, and L.~Slepyan, ``Analysis of dynamic damage
  propagation in discrete beam structures,'' {\em International Journal of
  Solids and Structures}, vol.~97, pp.~699--713, 2016.

\bibitem{thomson1971lattice}
R.~Thomson, C.~Hsieh, and V.~Rana, ``Lattice trapping of fracture cracks,''
  {\em Journal of Applied Physics}, vol.~42, no.~8, pp.~3154--3160, 1971.

\bibitem{gorbushin2017dynamic}
N.~Gorbushin and G.~Mishuris, ``Dynamic fracture of a discrete media under
  moving load,'' {\em Accepted for publication of International Journal of
  Solids and Structures. preprint arXiv:1701.02725}, 2017.

\bibitem{Alves2017}
L.~M. Alves and R.~F. R.~M. Lobo, ``The possibility to predict crack patterns
  on dynamic fracture,'' {\em International Journal of Fracture}, pp.~1--23,
  2017.

\bibitem{mishuris2008dynamics}
G.~S. Mishuris, A.~B. Movchan, L.~I. Slepyan, {\em et~al.}, ``Dynamics of a
  bridged crack in a discrete lattice,'' {\em Quarterly journal of mechanics
  and applied mathematics}, vol.~61, no.~2, pp.~151--160, 2008.

\bibitem{mishuris2009localised}
G.~S. Mishuris, A.~B. Movchan, and L.~I. Slepyan, ``Localised knife waves in a
  structured interface,'' {\em Journal of the Mechanics and Physics of Solids},
  vol.~57, no.~12, pp.~1958--1979, 2009.

\bibitem{slepyan2010crack}
L.~I. Slepyan, A.~B. Movchan, and G.~S. Mishuris, ``Crack in a lattice
  waveguide,'' {\em International Journal of Fracture}, vol.~162, no.~1,
  pp.~91--106, 2010.

\bibitem{slepyan2015forerunning}
L.~Slepyan, M.~Ayzenberg-Stepanenko, and G.~Mishuris, ``Forerunning mode
  transition in a continuous waveguide,'' {\em Journal of the Mechanics and
  Physics of Solids}, vol.~78, pp.~32--45, 2015.

\bibitem{nieves2017transient}
M.~Nieves, G.~Mishuris, and L.~Slepyan, ``Transient wave in a transformable
  periodic flexural structure,'' {\em International Journal of Solids and
  Structures}, vol.~112, pp.~185--208, 2017.

\bibitem{gorbushin2016cer}
N.~Gorbushin and G.~Mishuris, ``Dynamic crack propagation along the interface
  with non-local interactions,'' {\em Journal of the European Ceramic Society},
  vol.~36, no.~9, pp.~2241--2244, 2016.

\bibitem{ayzenberg2014brittle}
M.~Ayzenberg-Stepanenko, G.~Mishuris, and L.~Slepyan, ``Brittle fracture in a
  periodic structure with internal potential energy. spontaneous crack
  propagation,'' in {\em Proc. R. Soc. A}, vol.~470, p.~20140121, The Royal
  Society, 2014.

\bibitem{Mishuris_Slepyan2007}
G.~Mishuris, A.~Movchan, and L.~Slepyan, ``Waves and fracture in an
  inhomogeneous lattice structure,'' {\em Waves in Random and Complex Media},
  vol.~17, no.~4, pp.~409--428, 2007.

\bibitem{Nieves_Mish2013}
M.~Nieves, A.~Movchan, I.~Jones, and G.~Mishuris, ``Propagation of slepyan's
  crack in a non-uniform elastic lattice,'' {\em Journal of the Mechanics and
  Physics of Solids}, vol.~61, no.~6, pp.~1464--1488, 2013.

\bibitem{xiao2009fracture}
J.~Xiao, J.~Staniszewski, and J.~Gillespie, ``Fracture and progressive failure
  of defective graphene sheets and carbon nanotubes,'' {\em Composite
  structures}, vol.~88, no.~4, pp.~602--609, 2009.

\bibitem{merkel1999energy}
R.~Merkel, P.~Nassoy, A.~Leung, K.~Ritchie, and E.~Evans, ``Energy landscapes
  of receptor--ligand bonds explored with dynamic force spectroscopy,'' {\em
  Nature}, vol.~397, no.~6714, pp.~50--53, 1999.

\bibitem{hinterdorfer2006detection}
P.~Hinterdorfer and Y.~F. Dufr{\^e}ne, ``Detection and localization of single
  molecular recognition events using atomic force microscopy,'' {\em Nature
  methods}, vol.~3, no.~5, pp.~347--355, 2006.

\bibitem{Wei1998}
Y.~Wei and J.~Hutchinson, ``Interface strength, work of adhesion and plasticity
  in the peel test,'' {\em International Journal of Fracture}, vol.~1,
  p.~315–333, 1998.

\bibitem{Nase_Eremeev2016}
M.~Nase, M.~Rennert, K.~Naumenko, and V.~A. Eremeyev, ``Identifying
  traction–separation behavior of self-adhesive polymeric films from in situ
  digital images under t-peeling,'' {\em Int. J. Fract.}, vol.~91, pp.~40--55,
  2016.

\end{thebibliography}
\end{document}